# Appearance of ferroelectricity in BaO nanowires

Anna N. Morozovska [*,1], E.A. Eliseev[2], Robert Blinc[3] and Maya D. Glinchuk[2,†]

[1]V. Lashkarev Institute of Semiconductor Physics, NAS of Ukraine,
41, pr. Nauki, 03028 Kiev, Ukraine

[2]Institute for Problems of Materials Science, NAS of Ukraine,
Krjijanovskogo 3, 03142 Kiev, Ukraine

[3]Jožef Stefan Institute, P. O. Box 3000, 1001 Ljubljana, Slovenia

**Abstract**

We predict that ferroelectric phase can be induced by the strong intrinsic surface stress inevitably present under the curved surface in the high aspect ratio cylindrical nanoparticles of nonferroelectric binary oxides (BaO, EuO, MgO, etc).

We calculated the sizes and temperature range of the ferroelectric phase in BaO nanowires. The analytical calculations were performed within Landau-Ginzburg-Devonshire theory with phenomenological parameters extracted from the first principle calculations [E. Bousquet et al, Strain-induced ferroelectricity in simple rocksalt binary oxides. arXiv:0906.4235v1] and tabulated experimental data. In accordance with our calculations BaO nanowires of radius ~(1−10) nm can be ferroelectric at room temperature (with spontaneous polarization values up to 0.5 C/m$^2$) for the typical surface stress coefficients ~ (10−50) N/m.

We hope that our prediction can stimulate both experimental studies of rocksalt binary oxides nanoparticles polar properties as well as the first principle calculations of their spontaneous dipole moment induced by the intrinsic stress under the curved surface.

## I. Introduction

The substantial progress takes place in the last years in synthesis of various ferroics nanosystems [1], like ferroelectric thin films [2, 3], spherical nanoparticles with controllable sizes [4], nanotube-patterned films [5], aligned arrays of tubes [6, 7, 8, 9], wires [10] and rods with enhanced polar properties [11, 12]. These facts determine the interest to the nanosystems theoretical description. Then possibility to induce and govern the phase transitions in ferroic nanostructures due to the size and strain effects has been studied theoretically both from the first-principle microscopic theories [13, 14, 15] and Landau-Ginzburg-Devonshire (LGD) phenomenology for thin films [16, 17, 18, 19], nanospheres [20, 21], nanorods [22, 23, 24], nanowires [25] and nanotubes [26, 27].

---

[*] Corresponding author, morozo@i.com.ua;

[†] Corresponding author, glin@imps.kiev.ua

In particular enhanced ferroelectricity was observed in [11] and was shown to originate from surface tension phenomena [22] and appearance of ferroelectricity was predicted in thin strained films of incipient ferroelectrics $KTaO_3$ [28], as well as in epitaxially strained layers and superlattices of simple binary oxides BaO and EuO [29]. These intriguing predictions of new phenomena absent in bulk materials are waiting for experimental justification and would have numerous new applications.

However, it may happen that predicted strain-induced ferroelectricity would be hindered in many experimental cases for the BaO and especially EuO thin films on realistic substrates, since the epitaxial misfit strains more than 3-5 % are necessary for the ferroelectricity appearance in accordance with predictions [29]. Such giant strains should either exponentially relax via dislocations appearance [30], or the growth of continuous films surface appeared hardly possible, instead the incoherent growth of islands take place for high misfit strains more than 3-5 % [31]. Actually, it is well-known experimental fact that misfit dislocations originate in epitaxial films when their thickness is more than the critical thickness $h_d$ of dislocations appearance. The thickness $h_d$ strongly decreases with misfit strain $u_m$ increase. In accordance with the simplest but adequate models $h_d \sim 1/u_m$ in the wide range of $u_m$, and typically the critical thickness $h_d$ appeared not more than several lattice constants for misfit strains more than 3-5 % (see e.g. see Fig. 2b and estimations in Ref. [30]).

In any case the phase transitions considered in Refs. [28, 29, 32] belong to the group of the transitions induced by external forces, namely by the misfit strain between a film and substrate in the cases [28, 29] or dipole impurities in the incipient ferroelectrics [32]. It is obvious, that in the freestanding films or without impurities in the incipient ferroelectrics the ferroelectric phase should be eliminated contrary to conventional spontaneous polarization.

On the other hand the spatial confinement, as the characteristic feature that cannot be eliminated, leads to the strong enhancement of ferroelectric properties up to the critical size disappearance (as observed earlier in Rochelle salt nanorods [11, 12], predicted by first principles calculations [14] and phenomenological theory [22, 25]). Thus it is reasonable to suppose, that the size-induced ferroelectricity could be observed experimentally in BaO or EuO nanoparticles (nanowires, nanotubes or nanospheres) allowing for the intrinsic surface stress ("surface tension") that spontaneously exists under the curved surface. The stress σ strongly increases as $1/R$ with the particle radius $R$ decrease. Meanwhile to the best of our knowledge there are no published experimental data about misfit dislocations appearance in nanoparticles of radius 1-10 nm and so the effect cannot be decreased or even destroyed by them.

Let us considered schematically the physical picture of spontaneous polarization appearance in the nanoparticles of nonferroelectric binary oxides. The possible atomic structure of the MO wire cross-section is shown in Figs. 1a. It is seen from the figure that the surface stress leads to the strains and corresponding bond length contraction in radial direction ρ and its elongation along the wire axes *z*. The surface stress-induced effect facilitates out-of plane displacement of the light $O^{-2}$ anions and thus can cause the spontaneous coherent symmetry breaking. As a result, the macroscopic dipole moment can appear in the longitudinal direction *z* (see Figs. 1b). Similar mechanism was proposed by us earlier [22] to explain the ferroelectric transition temperature increase in Rochelle salt nanorods [11, 12].

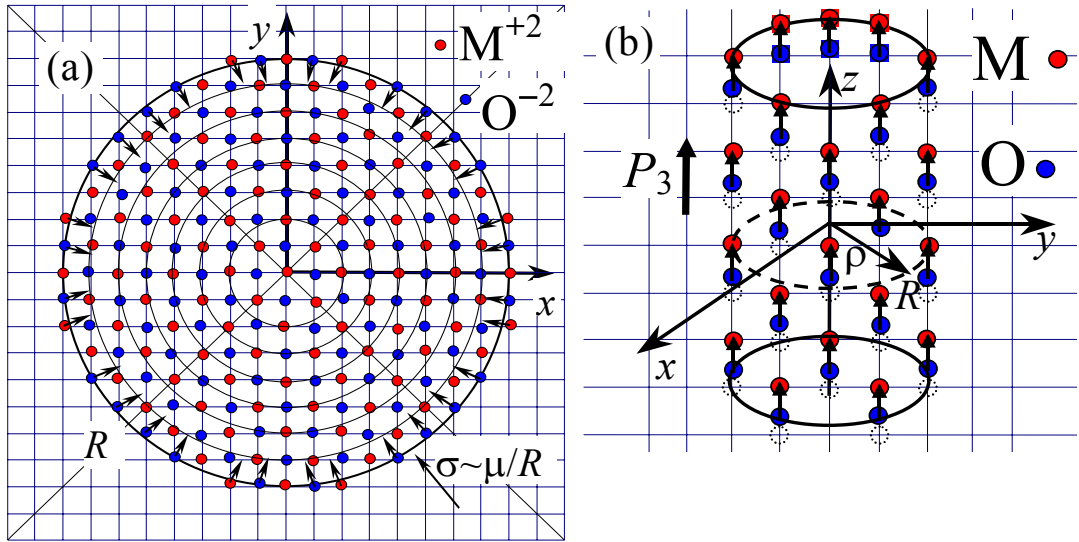

**Fig. 1**. Possible atomic structure of the MO wire: cross-section (a) and vertical section (b). Metal M is for Ba, Zn, Mg, Eu, etc.

In phenomenological approach used in this paper, the strains induced by the surface stress lead to the spontaneous polarization appearance via electrostriction, since the tetragonal symmetry distortions of unit cells facilitate polar ordering in non-ferroelectric media (see details in Ref. [22, 25]). To be sure that the spontaneous polarization and ferroelectric transition temperature can be high enough for observation at some reasonable values of nanoparticles radii it is necessary to calculate their values with the help of LGD theory with the help phenomenological parameters calculated for the bulk nonferroelectric system. It appeared possible to determine these parameterss for BaO only. Because of this below we calculate the sizes and temperature range of the ferroelectric phase induced in BaO nanowires by the surface curvature and size effects. Calculations are performed within LGD theory with phenomenological parameters extracted from the first principle calculations [29] for and tabulated experimental data.

## II. Basic equations

For correct phenomenological description of any nanosized system the surface energy should be considered, at that its contribution increases with the system size decrease [16-25]. Including the surface energy term, LGD free energy $F$ depends on the spontaneous polarization component $P_3$ and mechanical strains $u_{ij}$ as [17-18, 25]:

$$F = \begin{pmatrix} \int_V d^3r \left( \frac{\alpha}{2} P_3^2 + \frac{\beta}{4} P_3^4 + \frac{\gamma}{6} P_3^6 + \frac{g}{2}(\nabla P_3)^2 - P_3\left(E_e + \frac{E_3^d}{2}\right) - q_{ij33} u_{ij} P_3^2 + \frac{c_{ijkl}}{2} u_{ij} u_{kl} \right) \\ + \int_S d^2r \left( \frac{\alpha_S}{2} P_3^2 + \mu_{ij}^S u_{ij} \right) \end{pmatrix} \quad (1)$$

The surface energy coefficient $\alpha_S$ is regarded positive, isotropic and weekly temperature dependent, thus higher terms can be neglected in the surface energy expansion. Integration is performed over the system surface $S$ and volume $V$ correspondingly. $\mu_{ij}^S$ is the intrinsic surface stress tensor. The intrinsic surface stress exists under the curved surface of solid body and determines the excess pressure on the surface [33].

Expansion coefficient $\gamma > 0$ and the gradient term $g > 0$. Coefficient $\alpha(T) = \alpha_T(T - T_c^*)$, where $T$ is the absolute temperature. Value $T_c^*$ is positive Curie temperature for conventional ferroelectric bulk material with perovskite structure, while it is negative characteristic constant determined from the critical strain value at zero temperature for considered nonferroelectric BaO oxide layers with cubic but nonperovskite structure.

Stiffness tensor is $c_{ijkl}$; $q_{ijkl}$ stands for electrostriction stress tensor. $E_e$ is the external electric field. Considering long cylindrical nanoparticles with spontaneous polarization directed along the cylinder axes one could neglect the depolarization field $E_d$. At the same time the strong depolarization field existed in the spherical nanoparticles [21, 25].

For cubic symmetry nonzero components of the strain tensor inside cylindrical wire of the radius $R$ have the following form [25]: radial strain $u_{\rho\rho} = u_{11} = u_{22} = -(s_{11} + s_{12})(\mu/R) + Q_{12} P_3^2$ and longitudinal strain $u_{33} = -s_{12}(\mu/R) + Q_{11} P_3^2$, where $Q_{ij}$ are the components of electrostriction strain tensor, $s_{ij}$ are the elastic compliances, in the isotropic approximation $\mu_{ii}^S = \mu$. Note that the radial strain $u_{\rho\rho}$ is negative, while the longitudinal strain $u_{33}$ is positive at $P_3 = 0$, since $(s_{11} + s_{12}) > 0$ and $s_{12} < 0$. Thus the surface stress induces the bond lengths contraction in radial directions $\{x, y\}$ and their elongation in z direction. Corresponding tetragonality of the unit cell acquires the form:

$$\frac{c}{a} = \frac{1 + u_{33}}{1 + u_{22}} = \frac{1 - s_{12}(\mu/R) + Q_{11}\overline{P}^2}{1 - (s_{11} + s_{12})(\mu/R) + Q_{12}\overline{P}^2} \approx 1 + s_{11}\frac{\mu}{R} + (Q_{11} - Q_{12})\overline{P}^2. \quad (2)$$

Using the expressions for strains and direct variational method, the transition temperature into the ferroelectric phase and the spontaneous polarization $\bar{P}_3(R)$ averaged over the wire radius $R$ was derived similarly to [25] as:

$$T_{cr}(R) \approx T_c^* - Q_{12}\frac{4\mu}{\alpha_T R} - \frac{2}{\alpha_T}\left(\frac{g}{R\lambda + 2R^2/k_0^2}\right), \quad (3)$$

$$\bar{P}_3(R) \approx \sqrt{\frac{2\alpha_T(T_{cr}(R) - T)}{\beta + \sqrt{\beta^2 + 4\gamma\alpha_T(T_{cr}(R) - T)}}} \quad . \quad (4)$$

The constant $k_0 = 2.408...$ is the minimal root of Bessel function $J_0(k)$, $\lambda = g/\alpha_S$ is the so-called extrapolation length that depends on the surface energy contribution into the free energy.

The first negative size-independent term in Eq.(3) is determined from the critical strain value at zero temperature for nonferroelectric BaO oxide layers with flat surface, the second one is the contribution of intrinsic surface stress $\sim \mu/R$, the third negative term $\sim g$ originated from correlation effects. The third term decreases the possible transition temperature $T_{cr}$ at positive $\lambda$, while the second term increases $T_{cr}$ under the condition $Q_{12}\mu < 0$. By changing the wire radius one can tune the transition temperature and polarization value in the wide range.

LGD free energy expansion coefficients for BaO bulk material were extracted from the results of the first principle calculation [29] and experimental data [34] as described in Appendix. The surface tension coefficient $\mu$ varies in the typical range of 5–50 N/m [35] depending on the nanowire ambient (template material, gel or gas). All parameters are listed in the Tab.1.

**Table 1.** LGD free energy expansion coefficients for BaO bulk material.

| $\alpha_T$, m/(F K) | $\beta$, m$^5$/(C$^2$F) | $\gamma$, m$^9$/(C$^4$F) | $T_c^*$, K | $Q_{11}$, m$^4$/C$^2$ | $Q_{12}$, m$^4$/C$^2$ | $s_{11}$, m$^2$/N | $s_{12}$, m$^2$/N |
|---|---|---|---|---|---|---|---|
| $6.4 \cdot 10^6$ | $1.0 \cdot 10^9$ | $4.4 \cdot 10^{11}$ | -226 | 0.50±0.05 | -0.22±0.3 | $10.23 \cdot 10^{-12}$ | $-2.91 \cdot 10^{-12}$ |

Since $Q_{12}$ appeared negative, the surface curvature can induce ferroelectricity in BaO nanowires. The phase diagrams of BaO wires in coordinates temperature-radius, temperature-surface stress coefficient and contour maps in coordinates surface stress coefficient-radius are shown in Figs. 2. The dependence of the spontaneous polarization of BaO nanowires on temperature, radius and its contour maps in coordinates surface stress coefficient-radius are shown in Figs. 3. The figures were calculated from Eqs. (2-4), with the parameters from Tab. 1 and the gradient coefficient $g = 2 \cdot 10^{-9}$ m$^3$/F. Plots (a-c) in Figs. 2-3 correspond to $\lambda \gg R$, plots (d-f) correspond to $\lambda = 0$.

It is clear from the Figs. 2-3 that BaO nanowires of radius ~(1–10) nm can be ferroelectric at room and even at higher temperatures (with spontaneous polarization values up to 0.5 C/m$^2$) for the surface stress coefficient $\mu \sim (10-50)$ N/m.

As anticipated from Eqs.(3-4), the limiting case $\lambda \gg R$ is much more preferable for the ferroelectricity appearance, so as it corresponds to higher transition temperature and spontaneous polarization values. Let us underline that the dependence of the transition temperature $T_{cr}$ on the wire radius $R$ is monotonic for the case $\lambda \gg R$: $T_{cr}$ decreases with $R$ increase (Figs. 2a). Actually, for the case $T_{cr}(R) \approx T_c^* - Q_{12}\dfrac{4\mu}{\alpha_T R}$ and so the critical radius of the ferroelectricity appearance is $R_{cr}(T) \approx \dfrac{4\mu Q_{12}}{\alpha_T(T_c^* - T)}$ at a given temperature $T$. For the limiting case the influence of the correlation effects are negligibly small in comparison with the surface stress contribution. For the case $\lambda \gg R$ corresponding spontaneous polarization monotonically decreases with the wire radius increase and disappears at $R = R_{cr}(0)$ (see Fig. 3b).

In the opposite limiting case $\lambda = 0$ the region of the ferroelectricity appearance is the smaler (compare Figs. 2a and d). For the case $\lambda = 0$ the dependence of the transition temperature $T_{cr}$ on the wire radius $R$ is not monotonic: ferroelectricity appeared at zero Kelvins at the radius $R_{min}$, then the transition temperature increases for the radius $R_{min} < R < R_{opt}$ and decreases up to zero at radiuses $R_{opt} < R < R_{max}$, where $R_{min} = \dfrac{2\mu Q_{12} - \sqrt{\alpha_T T_c^* g k_0^2 + 4\mu^2 Q_{12}^2}}{\alpha_T T_c^*}$, $R_{max} = \dfrac{2\mu Q_{12} + \sqrt{\alpha_T T_c^* g k_0^2 + 4\mu^2 Q_{12}^2}}{\alpha_T T_c^*}$ and $R_{opt} = -\dfrac{g k_0^2}{2\mu Q_{12}}$. The maximal temperature at which ferroelectricity can be observed is $T_{cr}(R_{opt}) = T_c^* + \dfrac{4\mu^2 Q_{12}^2}{\alpha_T g k_0^2}$. The optimal radius $R_{opt}$ decreases and $T_{cr}(R_{opt})$ increases with the surface stress coefficient $\mu$ increase as anticipated. For the case $\lambda = 0$ the spontaneous polarization is not monotonic: $P_3$ appeared at some radius $R_{min}(T)$, then increases for the radius $R_{min}(T) < R < R_{opt}(T)$ and further decreases up to zero at radiuses $R_{opt}(T) < R < R_{max}(T)$ (see Fig. 3e).

At fixed radius $R$ the transition temperature (3) linearly increases with the surface stress coefficient $\mu$ increase (see curves in Figs. 2b,e). The spontaneous polarization (4) also increases with the surface tension coefficient increase (compare different curves in Figs. 3b,e).

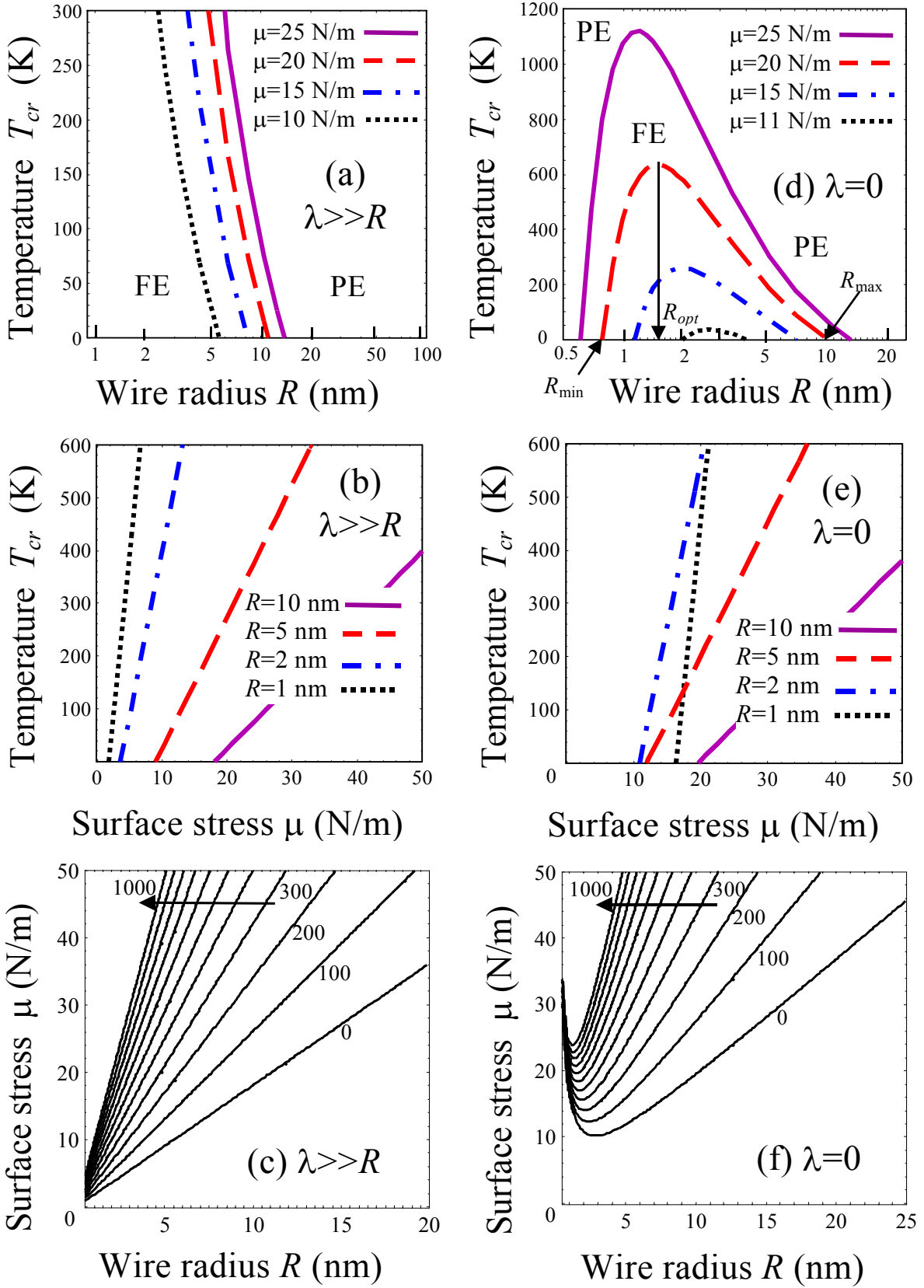

**Fig. 2**. Phase diagrams of BaO nanowires in coordinates temperature-radius (a,d), temperature-surface stress coefficient (b,e) and contour maps (c,f) in coordinates surface stress coefficient-radius. Figures near the contours (c,f) are $T_{cr}$ values in K. Plots (a-c) correspond to $\lambda \gg R$, plots (d-f) correspond to $\lambda = 0$.

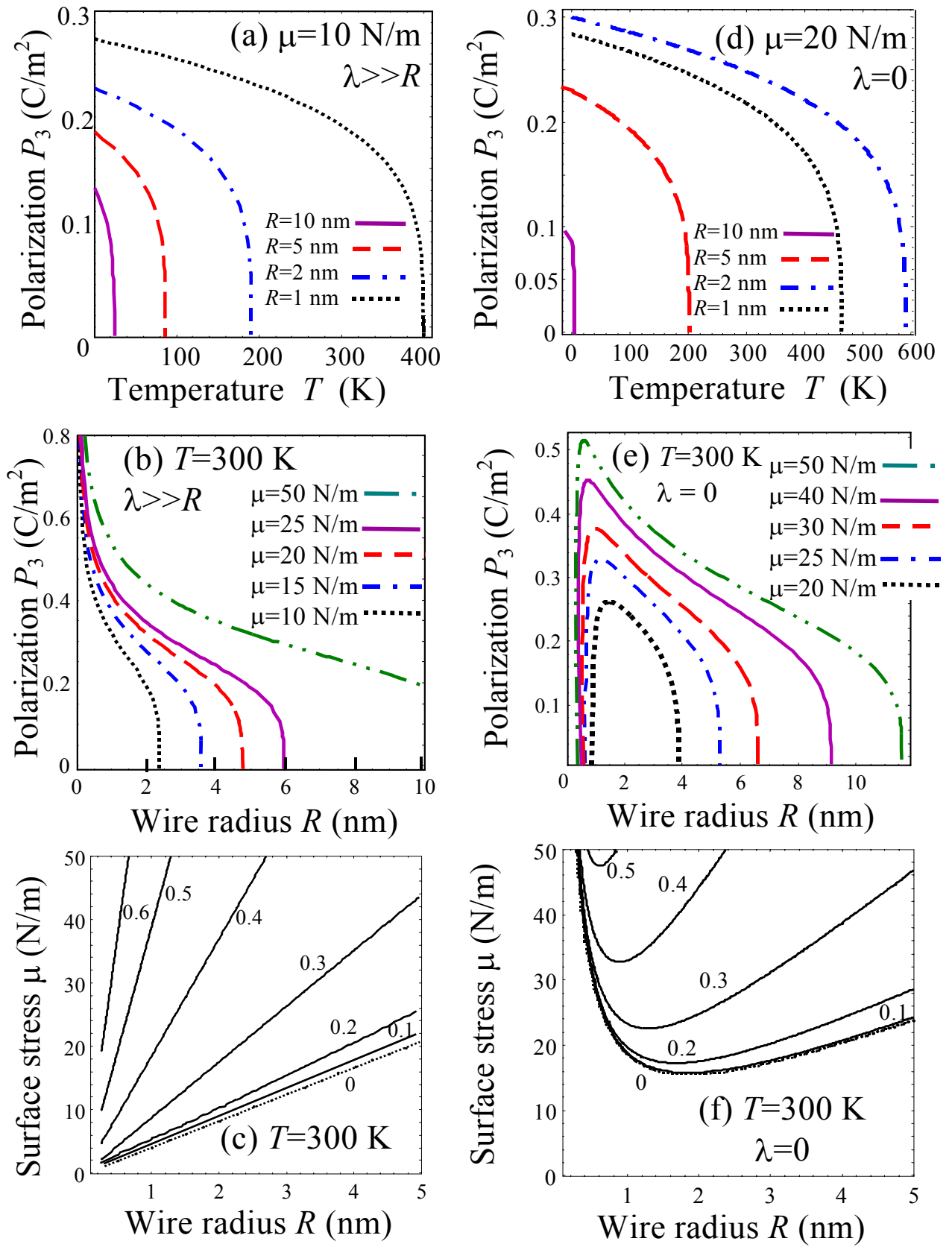

**Fig. 3**. Spontaneous polarization of BaO nanowires vs. temperature (a,d), radius (b,e) and contour maps (c,f) in coordinates surface stress coefficient-radius at $T$ = 300 K. Figures near the contours (c,f) are $P_3$ values in C/cm$^2$. Plots (a-c) correspond to $\lambda \gg R$, plots (d-f) correspond to $\lambda = 0$.

Contour maps shown in Figs. 2c,f and 3c,f allows one to optimize parameters and thus choose appropriate experimental conditions (wire radius, temperature interval and surface stress value) for the ferroelectricity observation in BaO nanowires.

Note, that similar calculations performed for the spherical BaO nanoparticles had shown that ferroelectricity in them could be hardly expected because of the uniform compressive strains $u_{11} = u_{22} = u_{33} = -(s_{11} + 2s_{12})(\mu/R)$ created by the surface stress and pretty strong depolarization field.

## Summary

We predict that ferroelectric phase can be induced by the strong intrinsic surface stress inevitably present under the curved surface and size effects in the high aspect ratio cylindrical nanoparticles of nonferroelectric binary oxides (BaO, EuO, MgO, etc).

Within the framework of LGD theory and using phenomenological parameters extracted from the first principle calculations [29] and tabulated experimental data we calculated the sizes and temperature range of the ferroelectric phase in BaO nanowires. In accordance with our calculations the nanowires of radius ∼(1−10) nm can be ferroelectric (with spontaneous polarization values up to 0.5 C/m$^2$) at room and even higher temperatures for the typical surface stress coefficients ∼ (10−50) N/m.

We hope that our prediction can stimulate both experimental studies of BaO nanoparticles polar properties as well as the first principle calculations.

## Acknowledgements

Research sponsored by National Academy of Sciences of Ukraine, Ministry of Science and Education of Ukraine (grant UU30/004) and National Science Foundation (Materials World Network, DMR-0908718).

**Appendix.**

**Calculations of the BaO material parameters**

BaO data from Ref. [34]: lattice constant: 0.5536 nm; linear thermal expansion coefficient: $13*10^{-6}$ $K^{-1}$; elastic stiffness components: $c_{11}$=126.1, $c_{12}$=50.0, $c_{44}$=33.7 (GPa). Then one could recalculate elastic compliance components via the relation $\hat{s} = \hat{c}^{-1}$ as $s_{11}$=10.23, $s_{12}$= – 2.91, $s_{44}$=29.67 ($10^{-12}$ $Pa^{-1}$).

Fitting with phenomenological theory the dielectric permittivity $\varepsilon_{33}$ component and spontaneous polarization dependence on misfit strain $u_m$ obtained by *ab-initio* calculations [29] for tetragonal phase one could estimate most of the necessary parameters. Below we assumed that DFT calculations [29] were performed at $T = 0$.

1) Using the relationship for paraelectric phase $(\varepsilon_{33} - 1)^{-1} = \varepsilon_0 (\alpha_T (T - T_c^*) - 4 u_m Q_{12}/(s_{11} + s_{12}))$, electrostriction constant could be estimated as $Q_{12} \approx -0.22$ $m^4/C^2$.

2) Using the permittivity $\varepsilon_{33}$ value 78 for zero misfit strain and zero temperature (from ab-initio calculations [29]) and value $\varepsilon_{33}$=34 at room temperature (from experiment data reviewed in Ref. [13]), one could get the following estimations: $\alpha_T \approx 6.4 \times 10^6$ $K^{-1}$ and $T_c \approx -226$ K

3) Fitting the misfit dependence for ferroelectric tetragonal phase, one could estimate nonlinear coefficients as $\beta \sim 10^9$, $\gamma = 4.4 \times 10^{11}$ SI units.

Other electrostriction coefficients could be obtained from piezoelectric strain coefficients $\hat{d} = \hat{e} \cdot \hat{s}$ using piezoelectric strain coefficients $\hat{e}$ calculated for BaO films in [34]. They are summarized in the Tab.S1.

**Table. S1**

| Piezo-strain (pm/V) | Misfit strain (%) | | | |
|---|---|---|---|---|
| | -2.67 | -1.63 | 1.99 | 3.03 |
| $d_{11}$ | - | - | 79.5 | 32.2 |
| $d_{31}=d_{32}$ | -16.3 | -33.4 | - | - |
| $d_{12}$ | - | - | 78.1 | 32.9 |
| $d_{13}$ | - | - | -116.2 | -47.0 |
| $d_{33}$ | 38.2 | 75.9 | - | - |
| $d_{24}=d_{15}$ | -8.6 | -5.0 | - | - |
| $d_{35}$ | - | - | -1.5 | -4.15 |
| $d_{26}$ | - | - | 416.9 | 157.2 |

So, using the relationships for tetragonal phase $d_{31} = 2\varepsilon_0\varepsilon_{33}Q_{12}P_3$, $d_{33} = 2\varepsilon_0\varepsilon_{33}Q_{11}P_3$ and polarization and dielectric permittivity calculated by Bousquet et al. one could estimate electrostriction coefficients for BaO as $Q_{12} \approx -(0.19 - 0.26)\,\text{m}^4/\text{C}^2$, $Q_{11} \approx 0.45 - 0.59\,\text{m}^4/\text{C}^2$. The discrepancy is related to the different sets of piezoelectric constants related to misfit strains -2.67 % and -1.63 %. Note, that independent data on dielectric permittivity dependence on misfit strain gives $Q_{12} \approx -0.22\,\text{m}^4/\text{C}^2$. Also the obtained electrostriction coefficients are consistent with tetragonality dependence on misfit strain, calculated by Bousquet et al.[29]

**References**

1 Scott JF (2007) Data storage: Multiferroic memories. Nature Materials **6**: 256-257

2 Fong DD, Stephenson GB, Streiffer SK, Eastman JA, Auciello O, Fuoss PH, Thompson C (2004) Ferroelectricity in Ultrathin Perovskite Films. *Science* **304**: 1650-1653

3 D.D. Fong, A.M. Kolpak, J.A. Eastman, S.K. Streiffer, P. H. Fuoss, G.B. Stephenson, Carol Thompson, D. M. Kim, K. J. Choi, C. B. Eom, I. Grinberg, and A. M. Rappe. Stabilization of Monodomain Polarization in Ultrathin PbTiO$_3$ Films. Phys. Rev. Lett. **96**, 127601 (2006).

4 Erdem E, Semmelhack H-C, Bottcher R, Rumpf H, Banys J, Matthes A, Glasel H-J, Hirsch D, Hartmann E (2006) Study of the tetragonal-to-cubic phase transition in PbTiO3 nanopowders. *J Phys: Condens Matter* **18**: 3861–3874

5 Poyato R and Huey BD and Padture NP (2006) Local piezoelectric and ferroelectric responses in nanotube-patterned thin films of BaTiO$_3$ synthesized hydrothermally at 200 °C *J Mater Res* **21**: 547-551.

6 Luo Y, Szafraniak I, Zakharov ND, Nagarajan V, Steinhart M, Wehrspohn RB, Wendorff JH, Ramesh R, Alexe M (2003) Nanoshell tubes of ferroelectric lead zirconate titanate and barium titanate. *Appl Phys Lett* **83**: 440-442

7 Morrison FD, Ramsay L, Scott JF (2003) High aspect ratio piezoelectric strontium-bismuth-tantalate nanotubes. *J Phys: Condens Matter* **15**: L527-L532.

8 Mishina ED, Vorotilov KA, Vasil'ev VA, Sigov AS, Ohta N, and Nakabayashi S. (2002) Porous silicon-based ferroelectric nanostructures. *J Exp Theor Phys* **95**: 502-504

9 Nourmohammadi A, Bahrevar MA, Schulze S and Hietschold M (2008) Electrodeposition of lead zirconate titanate nanotubes. *J Mater Sci* **43**: 4753–4759

10 Z.H. Zhou, X.S. Gao, J. Wang, K. Fujihara, and S. Ramakrishna, V. Nagarajan. Giant strain in PbZr$_{0.2}$Ti$_{0.8}$O$_3$ nanowires. Appl. Phys. Lett. **90**, 052902-1-3(2007)

11 D. Yadlovker, S. Berger. Uniform orientation and size of ferroelectric domains. Phys Rev B **71**: 184112-1-6 (2005).

12 D. Yadlovker, S. Berger. Ferroelectric single-crystall nano-rods grown within a nano-porous aluminium oxide matrix. Journal of Electroceramics, **22**, 281 (2009)

13 Ghosez P, Junquera J (2006) First-Principles Modeling of Ferroelectric Oxides Nanostructures. Chapter 134 of Handbook of Theoretical and Computational Nanotechnology Edited by Rieth M and Schommers W, American Scientifc Publisher, Stevenson Ranch

14 Geneste G, Bousquest E, Junquera J, Chosez P (2006) Finite-size effects in $BaTiO_3$ nanowires. *Appl Phys Lett* **88**: 112906-1-3

15 M.S. Majdoub, P. Sharma, T. Cagin, Dramatic enhancement in energy harvesting for a narrow range of dimensions in piezoelectric nanostructures. Phys Rev B **78**, 121407(R)-1-4 (2008).

16 Woo CH, and Zheng Y (2008) Depolarization in modeling nano-scale ferroelectrics using the Landau free energy functional. Applied Physics A: Materials Science & Processing 91: 59-63

17 Ban Z-G, Alpay SP, Mantese JV (2003) Fundamentals of graded ferroic materials and devices. *Phys Rev B* **67**: 184104-1-6

18 Akcay G, Alpay SP, Rossetti GA, and Scott JF (2008) Influence of mechanical boundary conditions on the electrocaloric properties of ferroelectric thin films *J Appl Phys* **103**: 024104-1-7

19 Qiu QY, Nagarajan V, Alpay SP (2008) Film Thickness-Misfit Strain Phase Diagrams for Epitaxial PbTiO3 Ultra-thin Ferroelectric Films. *Phys Rev B* **78**: 064117-1-13

20 Huang H, Sun CQ, Tianshu Zh, Hing P (2001) Grain-size effect on ferroelectric $Pb(Zr_{1-x}Ti_x)O_3$ solid solutions induced by surface bond contraction. *Phys Rev* B **63**: 184112-1-9

21 Glinchuk MD, Morozovska AN. (2003) Effect of Surface Tension and Depolarization Field on Ferroelectric Nanomaterials Properties. *Phys. Stat. Sol (*b) **238**: 81-91

22 A.N. Morozovska, E.A. Eliseev, and M.D. Glinchuk, Ferroelectricity enhancement in confined nanorods: Direct variational method. Phys. Rev. B **73:** 214106-1-13(2006)

23 Yue Zheng, C.H. Woo, and Biao Wang. Pulse-loaded ferroelectric nanowires as an alternating current source. Nano Letters **8,** 3131 (2008).

24 Yue Zheng, C.H. Woo, Biao Wang, Controlling dielectric and pyroelectric properties of compositionally graded ferroelectric rods by applied pressure. Journal of Applied Physics, **101**, 116103 (2007)

25 A.N. Morozovska, M.D. Glinchuk, E.A. Eliseev. Phase transitions induced by confinement of ferroic nanoparticles. Physical Review B. **76**, № 1, 014102 (2007).

26 Yue Zheng, C.H. Woo, and Biao Wang. Surface tension and size effects in ferroelectric nanotubes. J. Phys.: Condens. Matter **20,** 135216 (2008).

27 A.N. Morozovska, E.A. Eliseev, G.S. Svechnikov, and S.V. Kalinin. Pyroelectric response of ferroelectric nanoparticles: size effect and electric energy harvesting. http://arxiv.org/abs/0908.2311

28 E.A. Eliseev, M.D. Glinchuk, A.N. Morozovska. Appearance of ferroelectricity in thin films of incipient ferroelectric. Phys. stat. sol. (b) **244**, № 10, 3660–3672 (2007).

29 Eric Bousquet, Nicola Spaldin and Philippe Ghosez, Strain-induced ferroelectricity in simple rocksalt binary oxides. arXiv:0906.4235v1

30 J.S. Speck and W. Pompe. Domain configurations due to multiple misfit relaxation mechanisms in epitaxial ferroelectric thin films. I. Theory J. Appl. Phys. 76 (l), 466 (1994)

31 R.A. Rao, Q. Gan, C.B. Eom, Growth mechanisms of epitaxial metallic oxide SrRuO3 thin films studied by scanning tunneling microscopy. Appl. Phys. Lett. **71** (9), 1171 (1997)

32 B.E. Vugmeister, M.D. Glinchuk, Dipole glass and ferroelectricity in random −site electric dipole systems. Rev. Mod. Phys. **62**, 993 (1980).

33. V.A. Shchukin, D. Bimberg, Spontaneous ordering of nanostructures on crystal surfaces. Rev. Mod. Phys. **71**(4), 1125-1171 (1999).

34 Otfried Madelung, Semiconductors: Data Handbook, 3rd edition, Springer Verlag, Berlin (2004).

35. W. Ma, M. Zhang, Z. Lu. A study of size effects in $PbTiO_3$ nanocrystals by Raman spectroscopy Phys. Stat. Sol. (a). 166, № 2, 811-815 (1998).